\documentclass[3p]{elsarticle}

\usepackage{amssymb,amsmath,amsthm,epsfig,amsfonts,dsfont}
\usepackage[english]{babel}
\newcommand{\bX}{\boldsymbol{X}}
\newcommand{\bx}{\boldsymbol{x}}
\newcommand{\ba}{\boldsymbol{a}}
\newcommand{\bg}{\boldsymbol{g}}
\newcommand{\by}{\boldsymbol{y}}
\newcommand{\bz}{\boldsymbol{z}}
\newcommand{\bp}{\boldsymbol{p}}
\newcommand{\bs}{\boldsymbol{s}}
\newcommand{\f}{\boldsymbol{f}}
\newcommand{\bA}{\boldsymbol{A}}
\newcommand{\bB}{\boldsymbol{B}}
\newcommand{\bC}{\boldsymbol{C}}
\newcommand{\bD}{\boldsymbol{D}}

\newcommand{\bI}{\boldsymbol{I}}
\newcommand{\bP}{\boldsymbol{P}}
\newcommand{\bW}{\boldsymbol{W}}
\newcommand{\bR}{\boldsymbol{R}}
\newcommand{\R}{\mathds{R}}
\newcommand{\bQ}{\boldsymbol{Q}}
\newcommand{\bS}{\boldsymbol{S}}
\newcommand{\bu}{\boldsymbol{u}}
\newcommand{\bv}{\boldsymbol{v}}
\newcommand{\bU}{\boldsymbol{U}}
\newcommand{\bV}{\boldsymbol{V}}
\newcommand{\bh}{\boldsymbol{h}}
\newcommand{\bY}{\boldsymbol{Y}}
\newcommand{\bZ}{\boldsymbol{Z}}
\newcommand{\bL}{\boldsymbol{L}}
\newcommand{\diag}{\textrm{diag}}

\newtheorem{theorem}{Theorem}

\begin{document}

\begin{frontmatter}

\title{Amplitude and phase dynamics of noisy oscillators}


\author{Michele Bonnin}

\address{Department of Electronics and Telecommunications, Politecnico di Torino,\\
Corso Duca degli Abruzzi 24, I--10129 Turin, Italy\\
michele.bonnin@polito.it\\
Tel. +39 011 0904089}

\begin{abstract}
A description in terms of phase and amplitude variables is given, for nonlinear oscillators subject to white Gaussian noise described by It\^o stochastic differential equations. The stochastic differential equations derived for the amplitude and the phase are rigorous, and their validity is not limited to the weak noise limit. If the noise intensity is small, the equations can be efficiently solved using asymptotic expansions. Formulas for the expected angular frequency, expected oscillation amplitude and amplitude variance are derived using It\^o calculus.
\end{abstract}


\end{frontmatter}


\section{Introduction}\label{intro}

Nonlinear oscillations are ubiquitous in natural sciences and technology \cite{winfree1980,kuramoto1984,buzsaki2006}. An ideal oscillator would exhibit a perfectly periodic behavior, represented by a limit cycle in the state space. However, the output of actual oscillators is always corrupted by different types of disturbances, such as internal noise sources, thermal noise, interactions with the environment and with other systems. The autonomous nature of oscillators implies that any time shifted version of a solution is also a solution, therefore at least one direction exists along which perturbations are preserved. This explains why oscillators are very sensitive to noise, and introducing even small random perturbations into oscillators leads to dramatic changes in their frequency spectra and timing properties. This phenomenon, peculiar to oscillators, is known as phase noise or timing jitter. As a consequence, characterizing how noise affects the dynamics of oscillators is of paramount importance.

Phase models and phase reduction methods are powerful tools for analyzing the effect of perturbations on oscillators \cite{winfree1980,kuramoto1984,pikovsky2003,brown2004}. Phase models are based on the idea to describe the state of an oscillator using a phase and an amplitude functions. The phase represents the projection of the perturbed trajectory onto a reference orbit, e.g. an unperturbed limit cycle. The amplitude function, instead, measures the deviation of the perturbed orbit from the reference limit cycle. This picture can be further simplified under the hypothesis that the the limit cycle is asymptotically stable, and the noise is weak. With these assumptions, one expects that deviations from the limit cycle remain small, and that the amplitude function can be approximated by its unperturbed value. Then only the phase dynamics is retained, obtaining a phase reduced model.

In the last few years, phase models and phase reduced models have been extensively studied for the systematic investigation of the phase noise problem in oscillators  \cite{teramae2004,teramae2009,yoshimura2008,galan2009,goldobin2010,moehlis2014}. On the one hand, phase reduction method proved to be a valuable tool to derive simplified, mathematically tractable models, whose analysis has significantly increased our understanding of the influence of noise on oscillators. On the other hand, most of these models are of little help in analyzing practical oscillators. In fact, an explicit formula for the phase function can be found only for few, trivial, oscillators. In all other cases one must resort to numerical methods that are either approximate in nature, or unsuitable for oscillators of order higher than the second \cite{izhikevich2007,guillamon2009,suvak2011,bonnin2012}. The consequence is twofold. First, phase models cannot be used to study and design practical oscillators, and second, experimental and/or numerical results obtained from real world oscillators cannot be readily used to test the validity of phase models. Another potential problem concerns the model order reduction procedure. Amplitude fluctuations can be neglected only if the relaxation to the stable orbit is instantaneous, or at least, if it occurs on a time scale much shorter than the typical time for the phase dynamics. Such an assumption is often taken more for mathematical convenience than being physically plausible. Moreover, when dealing with noisy perturbations the stochastic nature of the processes should be taken into account. Appropriate correction term should be introduced before order reduction, to deal with the possible correlation between stochastic variables and noise increments \cite{yoshimura2008,bonnin2013,bonnin2014}.

This paper proposes possible solutions for the aforementioned problems. A novel amplitude and phase description for nonlinear oscillators subject to white Gaussian noise is presented. The method is based on the generalization of a classical technique \cite{hale1969} to the case of It\^o stochastic differential equations. Contrary to other derivations, the proposed amplitude and phase description does not rely on undetermined phase functions, and its validity is not necessarily limited to the case of weak noise. It is shown that using an appropriate basis, a partial decoupling between the amplitude and the phase dynamics can be obtained. The resulting equations represent the ideal starting point for the derivation of phase reduced models. It is also shown that under suitable simplifying hypothesis, the proposed model reduces to other models previously described in literature, i.e.  previously proposed models are special cases of our description, obtained applying different degrees of approximation. For the case of weak noise, the amplitude and phase equations can be analyzed using an asymptotic expansion method. The equations are recast as a sequence of linear stochastic differential equations, that can be solved iteratively. Instead to solve these equations, we exploit the linearity and the properties of It\^o integrals to obtain a statistical characterization for the noisy oscillators, finding analytical formulas for the expectation values of the angular frequency, the amplitude and the amplitude variance. As an example, the technique is applied to a noisy Stuart--Landau oscillator.

\section{Amplitude and phase dynamics of noisy oscillators}\label{amplitude-phase-dynamics}
Nonlinear oscillators subject to white Gaussian noise can be conveniently described by the stochastic differential equation (SDE)
\begin{equation}
d \bX_t = \ba(\bX_t) \, d t+ \varepsilon \bB(\bX_t) \, d \bW_t \label{eq1}
\end{equation}
where $\bX_t: \R \mapsto \R^n$ is a stochastic process describing the state of the oscillator, $\ba: \R^n \mapsto \R^n$ is a vector valued function that defines the oscillator dynamics, $\bB : \R^n \mapsto \R^{n,m}$ is a real valued $n \times m$ matrix, $\varepsilon$ is a parameter that measures the noise intensity and $\bW_t : \R \mapsto \R^m$ is a vector of Brownian motion components. We shall assume $\ba \in \mathcal{C}^k(\Omega \subseteq \R^n)$, with $k\ge 1$ and that $\bB$ satisfy a Lipschitz condition, to guarantee the existence and uniqueness of the solution \cite{oksendal03}.

Depending on the definition adopted for the stochastic integral, the SDE \eqref{eq1} can be interpreted following two main schemes: Stratonovich or It\^o. If Stratonovich interpretation is used, the amplitude and the phase equations can be derived in a straightforward way using the standard procedure described in \cite{hale1969}, since traditional calculus rules apply. However the analysis of the resulting equations gets more difficult because of the ``look in the future property'' of Stratonovich stochastic integral \cite{oksendal03}. By contrast It\^o interpretation does not suffer of the anticipating nature, but a new set of calculus rules, known as It\^o calculus, must be used. Therefore, we shall assume that \eqref{eq1} is an It\^o equation, and we shall use It\^o calculus to derive and to analyze the amplitude and phase equations.

For $\varepsilon = 0$ equation \eqref{eq1} reduces to the system of ordinary differential equations (ODEs)
\begin{equation}
\dfrac{d \bx(t)}{dt} = \ba(\bx(t)) \label{eq2}
\end{equation}
We assume that  \eqref{eq2} admits a $T$--periodic solution, i.e. a function $\bx_s(t)$ exists that satisfies \eqref{eq2} with the property $\bx_s(t + T) = \bx_s(t)$. Before proceed with the main result we introduce some notation. We define the unit vector tangent to the limit cycle (the symbol $'$ denotes the derivative with respect to the argument)
\begin{equation}
\bu_1(t) = \dfrac{\bx_s'(t)}{|\bx_s'(t)|} \label{eq3}
\end{equation}
Together with $\bu_1(t)$ we consider other $n-1$ linear independent vectors $\bu_2(t),\ldots,\bu_n(t)$, such that the set $\{\bu_1(t),\ldots,\bu_n(t)\}$ is a basis for $\R^n$, for all $t$. We remark that, differently from traditional derivations \cite{hale1969}, we require the vectors $\{\bu_1(t),\ldots,\bu_n(t)\}$ to be linearly independent, but not necessarily orthogonal. The conditions for the existence and a procedure to construct an orthogonal set are discussed, for instance, in \cite{hale1969,chow1998}. Together with $\{\bu_1(t),\ldots,\bu_n(t)\}$ we also consider another basis, $\{\bv_1(t),\ldots,\bv_n(t)\}$ constructed as follow: Given the matrix $\bU(t) = [\bu_1(t),\ldots,\bu_n(t)]$, we define the reciprocal vectors $\bv_1^T(t),\ldots,\bv_n^T(t)$ to be the rows of the inverse matrix $\bV(t) = \bU^{-1}(t)$. Thus $\{\bv_1(t),\ldots,\bv_n(t)\}$ also span $\R^n$ and the bi--orthogonality condition $\bv_i^T \bu_j = \bu_i^T \bv_j = \delta_{ij}$ holds. We shall also use the matrices $\bY(t) = [\bu_2(t),\ldots,\bu_n(t)]$, $\bZ(t) = [\bv_2(t),\ldots,\bv_n(t)]$, and the modulus of the vector field evaluated on the limit cycle, $r(t) = |\ba(\bx_s(t))|$.

The most important concept to be defined in the analysis of oscillator noise is the phase concept. A phase function is intended to represent the projection of the oscillator's state onto a reference trajectory, normally the unperturbed limit cycle. Being associated to a neutrally stable direction, random fluctuations of the phase will be neither adsorbed nor amplified, they persist and may eventually accumulate with time growing unboundedly large. This explains why the phase is a privileged variable.

We introduce a phase function $\theta : \R^n \mapsto [0,T)$, interpreted as an elapsed time from an initial reference point. Consider a point $\bx_s(0)$ on the limit cycle, and assign phase zero to this point, i.e. $\theta(\bx_s(0)) = 0$. The phase of the point $\bx_s(t)$ is $\theta(\bx_s(t)) = t, \mod T$. Thus, the phase represents a new parametrization of the limit cycle. Together with the phase function we shall consider an amplitude function $\bR : \R^n \mapsto \R^{n-1}$, with  $\theta, \bR \in \mathcal{C}^m(R^n)$, $m\ge2$. The amplitude\footnote{We shall use the term ``amplitude'' instead of the more correct ``amplitude deviation'' for the sake of simplicity.} function $\bR(\bx)$ is interpreted as an orbital deviation from the limit cycle. The following theorem represents the generalization of a classical result \cite{hale1969}, to the case of It\^o SDEs.

\begin{theorem}\label{theorem1}
Consider the It\^o SDEs \eqref{eq1} such that the ODEs obtained setting $\varepsilon=0$ admit a $T$--periodic limit cycle $\bx_s(t)$. Let $\{\bu_1(t),\ldots,\bu_n(t)\}$ and $\{\bv_1(t),\ldots,\bv_n(t)\}$ be two reciprocal bases such that $\bu_1(t)$ satisfies \eqref{eq3} and such that the bi--orthogonality condition $\bv_i^T \bu_j = \bu_i^T \bv_j = \delta_{ij}$ holds. Consider the coordinate transformation
\begin{equation}
\bx = \bh(\theta,\bR) = \bx_s(\theta(t)) + \bY(\theta(t)) \, \bR(t) \label{eq4}
\end{equation}
Then a neighborhood of the limit cycle $\bx_s(t)$ exists, where the phase $\theta(t)$ and the amplitude $\bR(t)$ are It\^o processes and satisfy
\begin{align}
d\theta = & \big[1 + a_1(\theta,\bR) + \varepsilon^2 \, \hat a_1(\theta,\bR) \big] dt + \varepsilon \bB_1(\theta,\bR) \, d\bW_t \label{eq5} \\[2ex]
d \bR = & \big[\bL(\theta) \bR+ \ba_2(\theta,\bR) + \varepsilon^2 \hat \ba_2(\theta,\bR) \big] dt + \varepsilon \bB_2(\theta,\bR) \, d\bW_t \label{eq6}
\end{align}
with (explicit dependence on $\theta$ and $t$ is omitted for simplicity)
\begin{align}
a_1(\theta,\bR)  = & \left(r + \bv_1^T\dfrac{\partial \bY}{\partial \theta} \bR \right)^{-1} \bv_1^T \bigg[\ba(\bx_s+\bY \bR) - \ba(\bx_s) -\dfrac{\partial \bY}{\partial \theta} \bR \bigg] \label{eq7} \\[1ex]
%%%%%%%%%%%%%%%%%%%%%%%%%%%%%%%%%%%%%%%%%%%%%%%%%%%%%%%%%%%%%%%%%%%%%%%%%%%%%%
\hat a_1(\theta,\bR) = &  - \left(r + \bv_1^T\dfrac{\partial \bY}{\partial \theta} \bR \right)^{-1} \bv_1^T  \bigg[ \dfrac{\partial \bY}{\partial \theta} \bB_2 \bB_1^T  + \dfrac{1}{2} \bigg( \dfrac{\partial \ba(\bx_s)}{\partial \theta} + \dfrac{\partial^2 \bY}{\partial \theta^2} \, \bR \bigg) \bB_1 \bB_1^T \bigg]\label{eq8} \\[1ex]
%%%%%%%%%%%%%%%%%%%%%%%%%%%%%%%%%%%%%%%%%%%%%%%%%%%%%%%%%%%%%%%%%%%%%%%%%%%%%%%
\bB_1(\theta,\bR) = & \bigg(r + \bv_1^T\dfrac{\partial \bY}{\partial \theta} \bR \bigg)^{-1}  \bv_1^T \, \bB(\bx_s + \bY \bR) \label{eq9} \\[1ex]
%%%%%%%%%%%%%%%%%%%%%%%%%%%%%%%%%%%%%%%%%%%%%%%%%%%%%%%%%%%%%%%%%%%%%%%%%%%%%%%%
\bL(\theta) = & - \bZ^T \dfrac{\partial \bY}{\partial \theta} \label{eq10} \\[1ex]
%%%%%%%%%%%%%%%%%%%%%%%%%%%%%%%%%%%%%%%%%%%%%%%%%%%%%%%%%%%%%%%%%%%%%%%%%%%%%%%%
\ba_2(\theta,\bR) = & - \bZ^T \bigg[ \dfrac{\partial \bY}{\partial \theta} \bR \, \ba_1 - \ba(\bx_s + \bY \bR) \bigg] \label{eq11} \\[1ex]
%%%%%%%%%%%%%%%%%%%%%%%%%%%%%%%%%%%%%%%%%%%%%%%%%%%%%%%%%%%%%%%%%%%%%%%%%%%%%%%%
\hat \ba_2(\theta,\bR)  = &  -\bZ^T \bigg[ \dfrac{\partial \bY}{\partial \theta} \bR \, \hat \ba_1 + \dfrac{1}{2} \bigg( \dfrac{\partial \ba(\bx_s)}{\partial \theta} + \dfrac{\partial^2 \bY}{\partial \theta^2} \, \bR \bigg) \bB_1 \bB_1^T  + \dfrac{\partial \bY}{\partial \theta} \bB_2 \bB_1^T \bigg] \label{eq12} \\[1ex]
%%%%%%%%%%%%%%%%%%%%%%%%%%%%%%%%%%%%%%%%%%%%%%%%%%%%%%%%%%%%%%%%%%%%%%%%%%%%%%%%
\bB_2(\theta,\bR) = & \bZ^T\bB(\bx_s+\bY \bR) - \bZ^T \dfrac{\partial \bY}{\partial \theta} \bR \, \bB_1(\bx_s + \bY \bR) \label{eq13}
\end{align}
\end{theorem}

\emph{Proof:}  First we show that a neighborhood of the limit cycle exists, where $\theta$ and $\bR$ are It\^o processes. To this end consider the coordinate transformation \eqref{eq4} and the associated Jacobian matrix
\begin{equation}
D\bh(\theta,\bR) = \left[\dfrac{\partial \bh}{\partial \theta}  \quad \dfrac{\partial \bh}{\partial \bR} \right] = \left[ \bx_s'(\theta) + \bY'(\theta) \bR \quad \bY(\theta) \right] \label{eq14}
\end{equation}
On the limit cycle $\bR=0$ and then
\[ D\bh(\theta,\bR)\big|_{\bR=0} = [\bx_s'(\theta) \quad \bY(\theta)] = [r(\theta)\bu_1(\theta), \bu_2(\theta),\ldots, \bu_n(\theta)] \]
Since $\{\bu_1(t),\ldots,\bu_n(t)\}$ is a basis for $\R^n$, it follows that the determinant of the Jacobian matrix is not null. Then by the Inverse Function Theorem there exists a neighborhood of $\bR=0$ where $\bh$ is invertible. Moreover, if $\bh$ is of class $\mathcal{C}^k$ then its inverse is also of class $\mathcal{C}^k$. Taking the inverse of $\bh$ we can write $\theta = \theta(\bx)$ and $\bR = \bR(\bx)$, and if the basis vectors are smooth enough it follows from  It\^o formula that $\theta$ and $\bR$ are It\^o processes.

Using It\^o formula and eq. \eqref{eq1}, $\bx = \bh(\theta,\bR)$ implies
\begin{align} \nonumber
d\bx = & \frac{\partial \bh}{\partial \theta} d\theta + \frac{\partial \bh}{\partial \bR} d\bR +  \frac{1}{2} \, \frac{\partial^2 \bh}{\partial \theta^2} (d\theta)^2 +  \frac{1}{2} \, d\bR^T \frac{\partial^2 \bh}{\partial \bR^2} d\bR
+  \frac{1}{2} \, \frac{\partial^2 \bh}{\partial \theta \partial \bR} d\theta d\bR \\[2ex]
= &  \ba(\bh(\theta,\bR)) dt + \varepsilon \bB(\bh(\theta,\bR)) \, d\bW_t \label{eq15}
\end{align}
where $\partial \bh/\partial \bR$ and $\partial^2 \bh/\partial \bR^2$ are the matrices of first and second partial derivatives with respect to the components of $\bR$, respectively. Introducing \eqref{eq4} in \eqref{eq15} yields
\begin{align} \nonumber
\bigg( \ba(\bx_s) + \dfrac{\partial \bY}{\partial \theta} \, \bR \bigg) d \theta + \bY d\bR + \dfrac{1}{2} \bigg( \dfrac{\partial \ba(\bx_s)}{\partial \theta} + \dfrac{\partial^2 \bY}{\partial \theta^2} \bR \bigg) (d\theta)^2 + \dfrac{\partial \bY}{\partial \theta} \, d\theta \, d\bR  \\
= \ba(\bx_s + \bY \bR) dt + \varepsilon \bB(\bx_s + \bY \bR) d\bW_t \label{eq16}
\end{align}
Multiplying to the left by $\bv_1^T$ and using the bi--orthogonality condition we get
\begin{align} \nonumber
\bigg( r + \bv_1^T \dfrac{\partial \bY}{\partial \theta} \, \bR \bigg) d \theta + \dfrac{1}{2} \bv_1^T \bigg( \dfrac{\partial \ba(\bx_s)}{\partial \theta} + \dfrac{\partial^2 \bY}{\partial \theta^2} \bR \bigg) (d\theta)^2
+ \bv_1^T \dfrac{\partial \bY}{\partial \theta} \, d\theta \, d\bR \\[1ex]
= \bv_1^T \ba(\bx_s+ \bY \bR) dt + \varepsilon \bv_1^T \bB(\bx_s+\bY \bR)  d\bW_t \label{eq17}
\end{align}
By converse, multiplying \eqref{eq16} to the left by $\bZ^T$ yields
\begin{align}\nonumber
\bZ^T \dfrac{\partial \bY}{\partial \theta} \bR \, d\theta + d\bR + \dfrac{1}{2} \bZ^T \bigg( \dfrac{\partial \ba(\bx_s)}{\partial \theta} + \dfrac{\partial^2 \bY}{\partial \theta^2} \bR \bigg) (d\theta)^2  + \bZ^T \dfrac{\partial \bY}{\partial \theta} \, d\theta \, d\bR \\[1ex]
= \bZ^T \ba(\bx_s + \bY \bR) dt + \varepsilon \bZ^T \bB(\bx_s + \bY \bR)  d\bW_t \label{eq18}
\end{align}
Since $\theta$ and $\bR$ are It\^o processes they satisfy relations of type $d \theta = \alpha \, dt + \varepsilon \boldsymbol{\beta} \, d\bW_t$, and $d \bR = \boldsymbol{\gamma} \, dt + \varepsilon \boldsymbol{\sigma} \, d\bW_t$, respectively. By It\^o lemma $(d\theta)^2 = \varepsilon^2 \boldsymbol{\beta}\boldsymbol{\beta}^T \, dt$,
$d\theta \, d\bR = \varepsilon^2  \boldsymbol{\sigma} \boldsymbol{\beta}^T \, dt$. Introducing these results in \eqref{eq17}, \eqref{eq18} and equating terms in $d\bW_t$ we obtain
\begin{align}
\boldsymbol{\beta} = & \left( r + \bv_1^T \frac{\partial \bY}{\partial \theta} \bR \right)^{-1} \bv_1^T \, \bB(\bx_s + \bY \, \bR) \label{eq19} \\[1ex]
\boldsymbol \sigma = & \bZ^T \bB(\bx_s + \bY \, \bR) - \bZ^T \frac{\partial \bY}{\partial \theta} \, \bR \,  \boldsymbol{\beta} \label{eq20}
\end{align}
Finally, using \eqref{eq19}, \eqref{eq20}, together with $(d\theta)^2 = \boldsymbol{\beta}\boldsymbol{\beta}^T \, dt$, $d\theta \, d\bR = \boldsymbol{\sigma} \boldsymbol{\beta}^T \, dt$ in \eqref{eq17}, $\eqref{eq18}$ and rearranging the terms we get the thesis. \hfill $\square$\\

\begin{figure}
\centering
 \includegraphics[width=67mm]{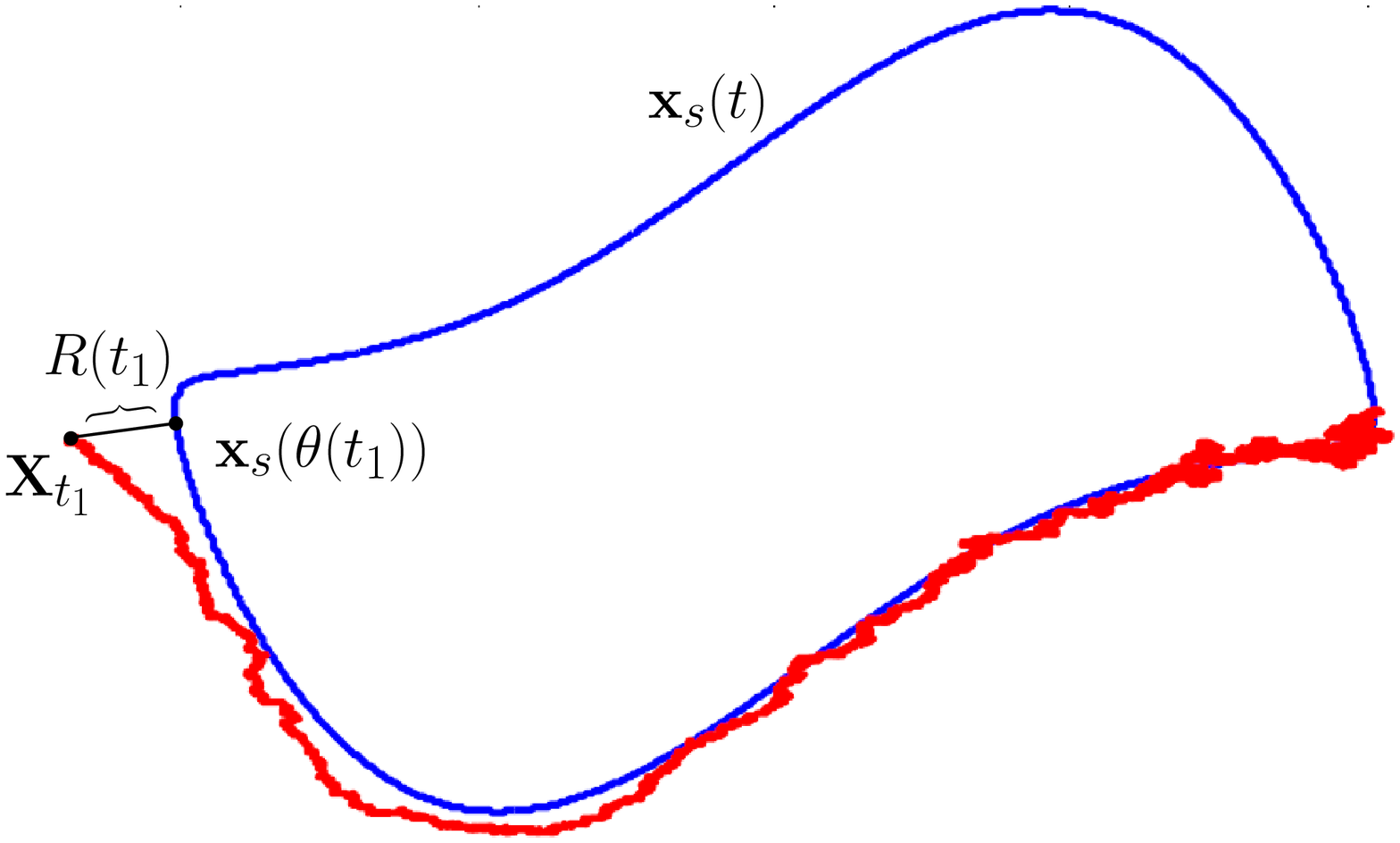}
 \includegraphics[width=67mm]{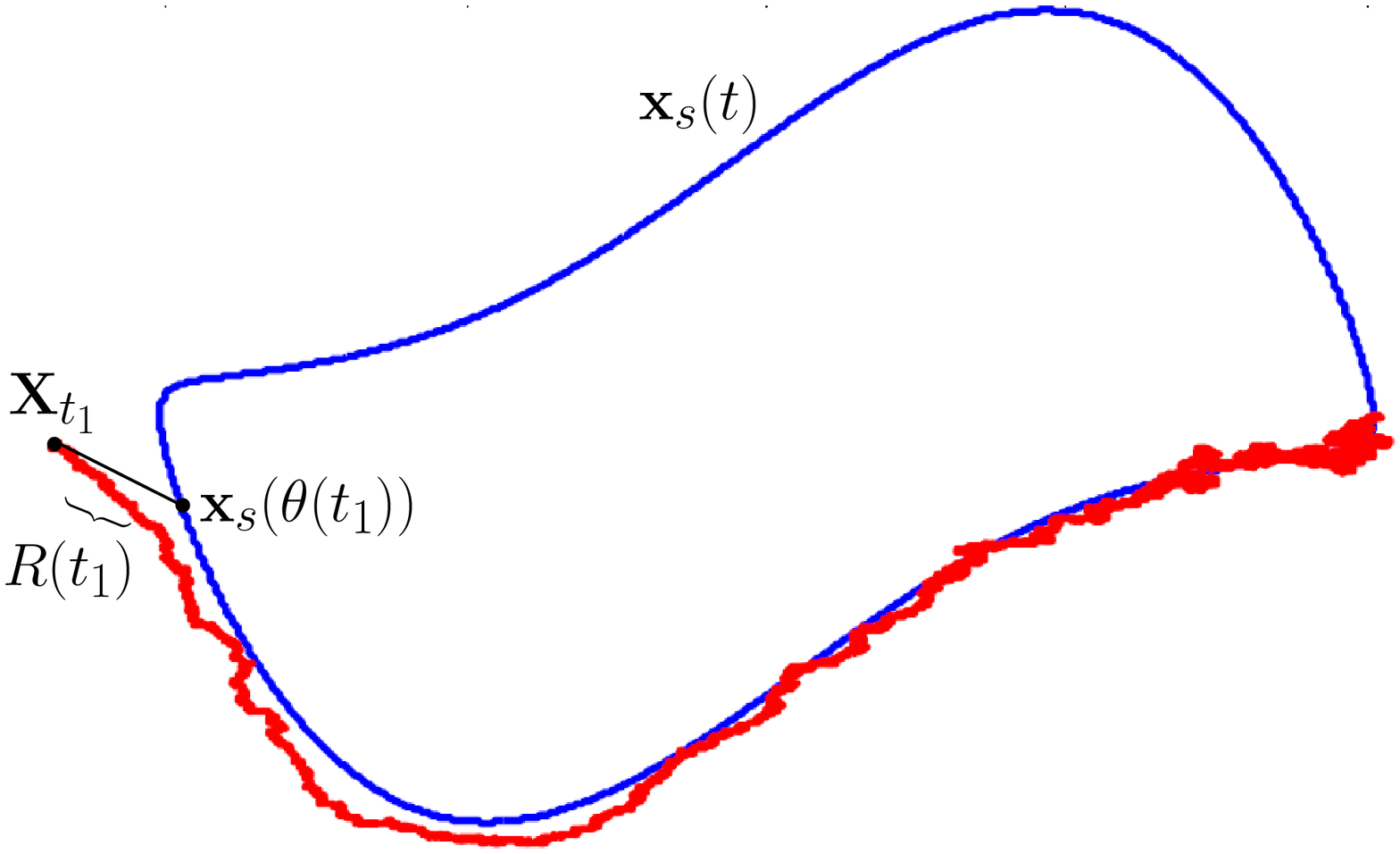}
\caption{Two possible decomposition of the stochastic process $\bX_t$. At the time $t_1$ the process is decomposed as $\bX_{t_1} = \bx_s(\theta(t_1)) + \bY(\theta(t_1)) R(t_1)$ using two different basis vectors. Left: orthogonal basis. Right: ``oblique'' basis. Red line is the stochastic process, blue line is the limit cycle shown for reference.\label{figure0}}
 \end{figure}
Figure \ref{figure0} shows the idea behind the decomposition defined by \eqref{eq4}. At any time instant $t_1$, the stochastic variable $\bX_{t_1}$ is split into two components. One component is tangent to the cycle, and it is given by the limit cycle evaluated at the stochastic time $\theta(t_1)$. The second component is transversal to the cycle, it is represented by the distance $\bR(t_1)$ measured along the vectors $\bu_2(\theta(t_1)),\ldots,\bu_n(\theta(t_1))$. Although the choice of an orthogonal frame may look the most natural, the use of particular non orthogonal frames
offers some advantages, as we shall show in the next section.

\section{Amplitude fluctuations and phase--amplitude decoupling}\label{decoupling}

The amplitude and phase equations \eqref{eq5} and \eqref{eq6} depend upon the choice of the basis vectors $\{\bu_1(t),\ldots,\bu_n(t)\}$. It is natural to ask whether a basis exists that has to be preferred to others. In particular, we shall show that, if the basis $\{\bu_1(t),\ldots,\bu_n(t)\}$ is chosen according to Floquet's theory, the phase dynamics can be partially decoupled from the amplitude dynamics. First we show that, under suitable conditions, the amplitude variable almost surely remains in a neighborhood of the unperturbed limit cycle. The following theorem represents an adaptation of a classical result \cite{hale1969}, to the case of bi--orthogonal basis.
\begin{theorem}\label{theorem2}
Consider the noiseless oscillator \eqref{eq2} with the $T$--periodic limit cycle $\bx_s(t)$. Let $1,\mu_2,\ldots,\mu_n$ be the characteristic multipliers of the variational equation
\begin{equation}
\dfrac{d\by(t)}{dt} = \bA(t) \, \by(t) \label{eq22}
\end{equation}
where $\bA(t) = \frac{\partial \ba(\bx_s(t))}{\partial \bx}$ is the Jacobian matrix evaluated over $\bx_s(t)$. Then $\bR=0$ is an equilibrium point for the amplitude equation with characteristic multipliers $\mu_2,\ldots,\mu_n$.
\end{theorem}
\emph{Proof:} For $\varepsilon=0$ the amplitude and phase equations \eqref{eq5}, \eqref{eq6}, reduce to
\begin{align}
\dfrac{d\theta}{dt} = & 1 + a_1(\theta,\bR) \label{eq23}\\[2ex]
\dfrac{d \bR}{dt} = & \bL(\theta) \bR+ \ba_2(\theta,\bR)  \label{eq24}
\end{align}
where $a_1(\theta,\bR)$, $\ba_2(\theta,\bR)$ and $\bL(\theta)$ are given by \eqref{eq7}, \eqref{eq10} and \eqref{eq11}, respectively. It is trivial to verify that $a_1(\theta,0)=0$ and $\ba_2(\theta,0)= 0$, thus $\bR=0$ is an equilibrium point for the amplitude equation \eqref{eq23}. The variational equation for the amplitude is obtained by taking the Taylor expansion of $\ba(\bx_s+\bY \bR)$ around $\bx_s(\theta)$ in the equation for $\ba_2(\theta,\bR)$ (eq. \eqref{eq11}). Using the fact that $\bZ^T \ba(\bx_s)= 0$ and neglecting $\mathcal{O}(\bR^2)$ terms we get
\begin{equation}
\dfrac{d\bR}{dt} = \left[\bL(\theta) + \bZ^T \bA\,\bY \right] \bR \label{eq25}
\end{equation}
Next we show that if $\bR_s(t)$ is a solution of \eqref{eq25}, then $\by_s(t) = \alpha \, \bx_s'(\theta) + \bY(\theta) \bR_s(t) $ with $\alpha \in \R$, is a solution of \eqref{eq22}. The following equality must hold
\begin{equation}
\dfrac{d\by_s}{d t} = \alpha \dfrac{\partial^2 \bx_s}{\partial \theta^2} + \dfrac{\partial \bY}{\partial \theta} \, \bR_s + \bY \dfrac{d \bR_s}{d t} = \bA \left( \alpha \dfrac{\partial \bx_s}{\partial \theta} + \bY \bR_s \right) \label{eq26}
\end{equation}
where \eqref{eq23} has been used and higher order terms in $\bR$ have been neglected.
It is well known that if $\bx_s$ solves \eqref{eq2}, then $\bx_s'$ solves \eqref{eq22}. Rearranging the terms, equation \eqref{eq26} reduces to
\begin{equation}
 \bY \dfrac{d \bR_s}{d t} = - \dfrac{\partial \bY}{\partial \theta} \bR_s +  \bA \, \bY \bR_s \label{eq27}
\end{equation}
Multiplying to the left for $\bZ^T$, and using the bi--orthogonality condition we obtain
\begin{equation}
\dfrac{d \bR_s}{d t} = \bZ^T \left( - \dfrac{\partial \bY}{\partial \theta} + \bA \bY \right) \bR_s \label{eq28}
\end{equation}
that coincides with \eqref{eq25}. Let $\bQ(t)$ be a fundamental matrix solution of the amplitude variational equation \eqref{eq25}, then a fundamental matrix solution of the variational equation \eqref{eq22} is of the form
\begin{equation}
\Phi(t) = \left[ \dfrac{\partial \bx_s}{\partial \theta},\;\dfrac{\partial \bx_s}{\partial \theta} [\alpha_1,\ldots,\alpha_{n-1}] + \bY \bQ(t) \right] \label{eq29}
\end{equation}
where $[\alpha_1,\ldots,\alpha_{n-1}]$ is a row vector of real constants. Since \eqref{eq22} is a linear system with $T$--periodic coefficients, there exists a constant matrix $\bC$ with eigenvalues $1,\mu_2,\ldots,\mu_n$, such that $\Phi(t+T)=\Phi(t) \bC$. From the periodicity of $\bx_s$ it follows that $\bC$ has the structure
\begin{equation}
\bC = \left[ \begin{array}{cc}
1 & \bC_2 \\
\mathbf{0} & \bC_1
\end{array} \right] \label{eq30}
\end{equation}
where the eigenvalues of $\bC_1$ are $\mu_2,\ldots,\mu_n$. On the other hand $\Phi(t+T)=\Phi(t) \bC$ together with the periodicity of $\theta$ imply
\begin{equation}
\dfrac{\partial \bx_s}{\partial \theta}[\alpha_1,\ldots,\alpha_{n-1}] +\bY \bQ(t+T) = \dfrac{\partial \bx_s}{\partial \theta}\left( \bC_2 + [\alpha_1,\ldots,\alpha_n]\bC_1\right) + \bY \bQ(t) \bC_1 \label{eq31}
\end{equation}
Multiplying to the left by $\bZ^T$ we obtain $\bQ(t+T) = \bQ(t) \bC_1$. Then the fundamental matrix solution of the amplitude variational equation \eqref{eq25} has multipliers $\mu_2,\ldots,\mu_n$ as required. \hfill$\square$

Theorem \ref{theorem2} implies that the amplitude equilibrium point $\bR=\mathbf{0}$ inherits the stability property of the limit cycle  $\bx_s(t)$, e.g. if $\bx_s(t)$ is asymptotically stable (i.e. $\mu_2,\ldots,\mu_n$ have modulus less than one), then $\bR = \mathbf{0}$ is also asymptotically stable. If this is the case, we expect that trajectories of the perturbed trajectory remains in a small neighborhood of the unperturbed limit cycle.

Since the amplitude is robust to perturbations, in most practical applications the phase is the only relevant variable. The following theorem establishes that using Floquet's theory, a particular basis can be identified such that a partial decoupling of the phase dynamics from the amplitude dynamics can be achieved. Before introducing the theorem we recall that from Floquet's theory, the fundamental matrix solution of the linear time periodic variational equation \eqref{eq22} can be written in the form
\begin{equation}
\Phi(t) = \bP(t) e^{\bD \,t} \bS_0 \label{eq32}
\end{equation}
where $\bP(t)$ is a $T$--periodic matrix, $\bS_0 = \bP^{-1}(0)$, and $\bD = \diag[\nu_1,\ldots,\nu_n]$ is a diagonal matrix whose diagonal entries are the Floquet's characteristic exponents \cite{farkas1994}.

%The fact that in presence of noise $\bR(t)$ remains small, suggests the possibility to derive an approximate phase reduced model, trading accuracy for simplicity. Perhaps, the simplest approach amounts in neglecting the amplitude fluctuations described by \eqref{eq6}, and substituting the stochastic process $\bR(t)$ with the constant unperturbed value $\bR=\mathbf{0}$ into \eqref{eq5} \cite{kuramoto1984,galan2009,yoshimura2008,teramae2009,moehlis2014}. The error introduced by this simplification depends not only on the magnitude of amplitude fluctuations, but also on the

\begin{theorem}\label{theorem3}
If the basis vectors $\{\bu_1(t),\ldots,\bu_n(t)\}$ are chosen such that $\big[ r(t) \bu_1(t),\bu_2(t),\ldots, \bu_n(t)\big] = \bP(t)$ then, the It\^o processes for the phase and amplitude reduce to
\begin{align}
d\theta  = & \big(1 + \widetilde a_1(\theta,\bR) + \varepsilon^2 \hat a_1(\theta,\bR)\big) dt + \varepsilon \bB_1(\theta,\bR) \, d\bW_t \label{eq33} \\[2ex]
d \bR = & \big( \widetilde \bD \, \bR + \widetilde \ba_2(\theta,\bR) + \varepsilon^2 \hat \ba_2(\theta,\bR) \big) d\, t + \varepsilon \bB_2(\theta,\bR) \, d\bW_t \label{eq34}
\end{align}
where $\widetilde \bD = \diag[\nu_2,\ldots,\nu_n]$, and the Taylor series of $\widetilde a_1(\theta,\bR)$, $\widetilde \ba_2(\theta,\bR)$, do not contain linear terms in $\bR$.
\end{theorem}
\emph{Proof:} It is well known from Floquet's theory that the columns of $\bP(t)$ are linearly independent for any $t$, and thus they can be chosen as a basis for $\R^n$. Moreover, $\bx_s'(t)$ is a solution for the variational equation, associated with the structural Floquet's exponent $\nu_1 = 0$. Thus we can take  the first column of $\bP(t)$ to be $r(t) \bu_1(t) = \bx_s'(t)$. Equation \eqref{eq32} implies $\bP(t) = \Phi(t) \bP(0) \, e^{-\bD \,t}$. Taking the derivative of \eqref{eq32} we have $\Phi'(t) = \bP' \, e^{\bD \,t} \bS_0 + \bP \, \bD \, e^{\bD \,t} \bS_0$, and taking into account that $\Phi(t)$ is a fundamental matrix of the variational equation \eqref{eq22} yields
$\bP' = \bA \bP - \bP \bD$. Removing the first column we get $\bY'= \bA \bY - \bY \widetilde \bD$. Substituting for $\bY'$ in \eqref{eq7}, \eqref{eq10} and \eqref{eq11}, taking the Taylor series $\ba(\bx_s + \bY \bR) = \ba(\bx_s) + \bA \bY \bR + \ldots$, and using the bi--orthogonality condition the thesis follows. \hfill $\square$

The partial decoupling established by theorem \ref{theorem3} together with the fact that, under the asymptotic stability hypothesis the amplitude fluctuations remain small, permit to derive simplified phase models. In particular, if amplitude fluctuations are very small, an adiabatic approximation $\bR(t)=\mathbf{0}$ may be used. If $\mathcal{O}(\varepsilon^2)$ terms are neglected, the phase dynamics reduces to
\begin{equation}
d\theta  = dt + \varepsilon \bB_1(\theta,\mathbf{0}) \, d\bW_t \label{eq35}
\end{equation}
that is equivalent to the traditional phase model derived in \cite{kuramoto1984,teramae2004}. Conversely, retaining $\mathcal{O}(\varepsilon^2)$ terms yields the phase equation
\begin{equation}
d\theta  =  \big(1 + \varepsilon^2 \hat a_1(\theta,\mathbf{0})\big) dt + \varepsilon \bB_1(\theta,\mathbf{0}) \, d\bW_t \label{eq36}
\end{equation}
analogous to those derived in \cite{yoshimura2008,bonnin2013,bonnin2014}.

\section{Small noise limit and asymptotic expansion}\label{weak-noise}

In general the phase and amplitude equations \eqref{eq5} and\eqref{eq6} are not easier to solve than the original SDE \eqref{eq1}. On the other hand, the phase reduced model \eqref{eq35} leads to incorrect predictions, such as that the noise does not influence the expected angular frequency of the oscillator \cite{bonnin2013}. Finally although it is more consistent from the point of view of physics, the reduced phase equation \eqref{eq36} may lead to inaccurate results as a consequence of the adiabatic approximation. In fact due to the nonlinear nature of oscillators, perturbations along certain directions are amplified, while other are reduced. The result is a net contribution to the amplitude deviation so that, contrary to the adiabatic assumption, the expected value of the amplitude deviation $E[\bR]$ is not null.

In what follows an alternative approach is proposed, which is based on the simultaneous solution of the phase and the amplitude equations. In the weak noise limit ($\varepsilon \ll1$) equations \eqref{eq5}, and \eqref{eq6}, can be efficiently solved using asymptotic expansions. For the sake of simplicity, we restrict the attention to second order oscillators, so that the amplitude deviation is a scalar variable. Higher order oscillators do not pose any particular problem, they only make the notation more involved. We search for solutions in the form $\theta = \theta_0 + \varepsilon \theta_1 + \varepsilon^2 \theta_2 + \ldots$, $R = R_0 + \varepsilon R_1 + \varepsilon^2 R_2 + \ldots$. Introducing these ansatzs in \eqref{eq5} and \eqref{eq6}, and equating the same powers of $\varepsilon$ we obtain:\\
Zeroth order
\begin{align}
d\theta_0 = &  \big[ 1 + a_1(\theta_0,R_0) \big] dt \label{eq37} \\[2ex]
dR_0  = & \big[ L(\theta_0) R_0 + a_2(\theta_0,R_0) \big] dt \label{eq38}
\end{align}
First order:
\begin{align}
d\theta_1 = &  \left(\frac{\partial a_1}{\partial \theta} \theta_1 +  \frac{\partial a_1}{\partial R} R_1\right) dt + \bB_1 d\bW_t \label{eq39} \\[1ex]
dR_1 = & \bigg[ \bigg(\frac{\partial L}{\partial \theta} R_0 + \frac{\partial a_2}{\partial \theta}\bigg) \theta_1 + \bigg(L + \frac{\partial a_2}{\partial R} \bigg) R_1 \bigg] dt + \bB_2 d\bW_t \label{eq40}
\end{align}
Second order:
\begin{align}
\nonumber d\theta_2  = &  \bigg[\frac{\partial a_1}{\partial \theta} \theta_2 +  \frac{\partial a_1}{\partial R} R_2 + \frac{1}{2} \frac{\partial^2  a_1}{\partial \theta^2} \theta_1^2 + \frac{1}{2} \frac{\partial^2  a_1}{\partial R^2} R_1^2 + \frac{\partial^2  a_1}{\partial \theta \partial R} \theta_1\, R_1 + \hat a_1 \bigg] dt \\[1ex]
& + \bigg[ \frac{\partial \bB_1}{\partial \theta} \theta_1 + \frac{\partial \bB_1}{\partial R} R_1  \bigg] d\bW_t \label{eq41} \\[2ex]
\nonumber dR_2  = & \bigg[ \bigg(\frac{\partial L}{\partial \theta} R_0 + \frac{\partial a_2}{\partial \theta}\bigg) \theta_2 + \bigg(L + \frac{\partial a_2}{\partial R} \bigg) R_2 + \frac{1}{2} \frac{\partial^2  a_2}{\partial R^2} R_1^2  \\
\nonumber & + \frac{1}{2} \left(\frac{\partial^2  L}{\partial \theta^2} R_0 + \frac{\partial^2  a_2}{\partial \theta^2} \right) \theta_1^2 + \bigg(\frac{\partial  L}{\partial \theta} + \frac{\partial^2  a_2}{\partial \theta \partial R} \bigg) \theta_1 \, R_1  + \hat a_2 \bigg] dt \\
 & + \bigg[ \frac{\partial \bB_2}{\partial \theta} \theta_2 + \frac{\partial \bB_2}{\partial R} R_2  \bigg] d\bW_t \label{eq42} \\
\nonumber & \vdots
\end{align}
In equations \eqref{eq39}--\eqref{eq42}, $L$ and its derivatives are evaluated at $\theta_0$, while $a_1$, $a_2$, $\hat a_1$, $\hat a_2$, $\bB_1$, $\bB_2$, and their derivatives are evaluated at $(\theta_0,R_0)$. The initial conditions are conveniently chosen in the form $\theta_n(0)=0$, $R_n(0) = $ for all $n$. The zeroth order equations \eqref{eq37}, \eqref{eq38} are nonlinear ODEs describing the noiseless oscillator. They have the simple solution $\theta_0=t$, $R_0=0$, that represents the limit cycle $\bx_s(t)$ in the phase and amplitude variables. Starting with equations \eqref{eq39}--\eqref{eq40}, we have a sequence of time dependent Ornstein--Uhlenbeck processes, whose analytical solution can be found using the method of integrating factors \cite{oksendal03}.  The first order solutions are introduced into second order equations, which can be solved by the same method. The process can be iterated to higher powers of $\varepsilon$ to obtain increasingly accurate solutions. The power series solution will not normally be a convergent series. However it is possible to show that the expansion is asymptotic, in the sense that the difference between the exact solution and its $\varepsilon^n$ order approximation is of order $\varepsilon^{n+1}$ \cite{gardiner1985}.

The explicit solution of equations \eqref{eq37}--\eqref{eq42} is of little use, because it depends on the particular realization of the stochastic process $\bW_t$, i.e. different realizations of the Brownian motion lead to different specific solutions. Nevertheless useful information can be obtained without solving the SDEs. Taking the expectation values on both sides of \eqref{eq39}--\eqref{eq42} and using the zero expectation property of It\^o integral, the following ODEs for the expectation values are found
\begin{align}
\frac{dE[\theta_1]}{dt}  = & E\left[ \frac{d\theta_1}{dt} \right] = \frac{\partial a_1}{\partial \theta} E[\theta_1] +  \frac{\partial a_1}{\partial R} E[R_1] \label{eq43}\\[1ex]
\frac{dE[R_1]}{dt} = & \bigg(\frac{\partial L}{\partial \theta} R_0 + \frac{\partial a_2}{\partial \theta}\bigg) E[\theta_1] + \bigg(L + \frac{\partial a_2}{\partial R} \bigg) E[R_1] \label{eq44}
\end{align}
\begin{align}
\nonumber \frac{dE[\theta_2]}{dt}  = & E\left[ \frac{d\theta_2}{dt} \right] = \frac{\partial a_1}{\partial \theta} E[\theta_2] +  \frac{\partial a_1}{\partial R} E[R_2] + \frac{1}{2} \frac{\partial^2  a_1}{\partial \theta^2} E[\theta_1^2] + \frac{1}{2} \frac{\partial^2  a_1}{\partial R^2} E[R_1^2] \\[1ex]
& + \frac{\partial^2  a_1}{\partial \theta \partial R} E[\theta_1\, R_1] + \hat a_1(\theta_0,R_0) \label{eq45}\\[2ex]
\nonumber \frac{dE[R_2]}{dt} = & \bigg(\frac{\partial L}{\partial \theta} R_0 + \frac{\partial a_2}{\partial \theta}\bigg) E[\theta_2] + \bigg(L + \frac{\partial a_2}{\partial R} \bigg) E[R_2]  + \frac{1}{2} \bigg(\frac{\partial^2  L}{\partial \theta^2} R_0  + \frac{\partial^2  a_2}{\partial \theta^2} \bigg) E[\theta_1^2] \\[1ex]
& + \frac{1}{2} \frac{\partial^2  a_2}{\partial R^2} E[R_1^2] + \big(\frac{\partial  L}{\partial \theta} + \frac{\partial^2  a_2}{\partial \theta \partial R} \big) E[\theta_1 \, R_1] + \hat a_2(\theta_0,R_0) \label{eq46}
\end{align}
Due to the linearity of the SDEs \eqref{eq39}--\eqref{eq42}, system \eqref{eq43}--\eqref{eq46} can be closed. Using It\^o formula, SDEs for $d(\theta_1^2)$, $d(R_1^2)$ and $d(\theta_1\,R_1)$ can be derived, and taking stochastic expectations the following additional equations are found
\begin{align}
\frac{dE[\theta_1^2]}{dt}  = & 2 \frac{\partial a_1}{\partial \theta} E[\theta_1^2] +  2 \frac{\partial a_1}{\partial R} E[\theta_1 R_1] + \bB_1 \bB_1^T \label{eq47}\\[1ex]
\frac{dE[R_1^2]}{dt}  = & 2 \big(L + \frac{\partial a_2}{\partial R} \big) E[R_1^2] + 2 \big(\frac{\partial L}{\partial \theta} R_0 + \frac{\partial a_2}{\partial \theta} \big) E[\theta_1 \, R_1] + \bB_2 \bB_2^T \label{eq48}\\[1ex]
\frac{dE[\theta_1 \, R_1]}{dt} = &  \big(L + \frac{\partial a_1}{\partial \theta} + \frac{\partial a_2}{\partial R} \big) E[\theta_1 \, R_1] + \frac{\partial a_1}{\partial R} E[R_1^2] + \big(\frac{\partial L}{\partial \theta} R_0 + \frac{\partial a_2}{\partial \theta} \big) E[\theta_1^2]  + \bB_1 \bB_2^T \label{eq49}
\end{align}
Equations \eqref{eq43}--\eqref{eq49} form a nonhomogeneous linear system of ordinary differential equations with time periodic coefficients. The equilibrium points allow to calculate the stationary expected angular frequency $E[d\theta/dt] = 1 + \varepsilon E[d\theta_1/dt] + \varepsilon^2 E[d\theta_2/dt]$, the expected amplitude $E[R] =  \varepsilon E[R_1] + \varepsilon^2 E[R_2]$ and amplitude variance $\textrm{var}(R) = \varepsilon^2 \textrm{var}(R_1) + \mathcal{O}(\varepsilon^3)$.

\section{Application}\label{example}
As an example of application, we consider a Stuart--Landau oscillator with multiplicative noise
\begin{equation}  \begin{array} {rcl}
d \varphi & = & \left( \alpha - \beta \rho^2 \right) dt + \varepsilon \, \rho \, d W_1 \\[1ex]
d \rho & = & \left(\rho - \rho^3 \right) dt + \varepsilon \, \rho^2 \, d W_2
\end{array} \label{eq50} \end{equation}
In absence of noise (for $\varepsilon =0$), the Stuart--Landau system admits an asymptotically stable limit cycle
\begin{equation}
\bx_s(t) = \left[ \begin{array}{c}
(\alpha-\beta)t \\
1 \end{array} \right] \label{eq51}
\end{equation}
so that the unit tangent vector is $\bu_1(t) = [1,0]^T$.

\subsection{Phase and amplitude equations with orthogonal basis}
Consider the orthogonal basis composed by $\bu_1(t)=[1,0]^T$, and $\bu_2(t)=[0,-1]^T$. Obviously $\bv_i(t)=\bu_i(t)$ for $i=1,2$. With this basis, the change of coordinates $\bx(t) = \bh(\theta,R)= \bx_s(\theta) + \bu_2(\theta) R$ implies $\varphi = (\alpha-\beta)\theta$ and $\rho = 1-R$. Using equations \eqref{eq7}--\eqref{eq13} it is straightforward to derive the phase and amplitude equations
\begin{equation} \begin{array}{rcl}
d\theta & = & \left(1+\dfrac{2\beta}{|\alpha - \beta|} (R-R^2)\right) dt + \dfrac{\varepsilon}{|\alpha -\beta|} \, (1-R) \, dW_1\\[2ex]
d R & = & \left( -2R+3R^2-R^3\right) dt - \varepsilon \, (1-R)^2 dW_2
\end{array} \label{eq52} \end{equation}
As expected in the equation for the phase a drift term linear in $R$ appears. The equations for the first few terms of the asymptotic expansion are
%\begin{equation}\begin{array}{rcl}
%d \theta_0 & = & \left[1+\dfrac{2\beta}{|\alpha - \beta|} \left(R_0 - R_0^2 \right) \right] dt \\[2ex]
%d R_0 & = & \left(-2 R_0 + 3 R_0^2 - R_0^3 \right) dt
%\end{array} \label{eq53} \end{equation}
%
%\begin{equation} \begin{array}{rcl}
%d \theta_1 & = & \dfrac{2\beta}{|\alpha - \beta|} \left(1 - 2 R_0  \right) R_1 \, dt + \dfrac{1}{|\alpha-\beta|} \big[(1-R_0) \, dW_1 \\[2ex]
%d R_1 & = & \left(-2 + 6 R_0 - 3 R_0^2 \right) R_1 \, dt - (1-R_0)^2 \, d W_2
%\end{array} \label{eq54} \end{equation}
%
%\begin{equation} \begin{array}{rcl}
%d \theta_2 & = & \dfrac{2\beta}{|\alpha - \beta|} \big[ \left(1 -2R_0\right) R_2 - R_1^2 \big]dt  - \dfrac{1}{|\alpha-\beta|} \, R_1 \, dW_1\\[2ex]
%d R_2 & = & \big[ \big(6 R_0 - 3 R_0^2 -2 \big) \, R_2 + 3 (1 - R_0) R_1^2 \big] \, dt + 2(1-R_0)\,R_1 \, d W_2
%\end{array} \label{eq55}  \end{equation}
\begin{align}
d \theta_0  = & \left[1+\dfrac{2\beta}{|\alpha - \beta|} \left(R_0 - R_0^2 \right) \right] dt \label{eq53}\\[2ex]
d R_0  = & \left(-2 R_0 + 3 R_0^2 - R_0^3 \right) dt \label{eq54}\\[2ex]
d \theta_1  = & \dfrac{2\beta}{|\alpha - \beta|} \left(1 - 2 R_0  \right) R_1 \, dt + \dfrac{1}{|\alpha-\beta|} \big[(1-R_0) \, dW_1 \label{eq55}\\[2ex]
d R_1  = & \left(-2 + 6 R_0 - 3 R_0^2 \right) R_1 \, dt - (1-R_0)^2 \, d W_2 \label{eq56}\\[2ex]
d \theta_2  = & \dfrac{2\beta}{|\alpha - \beta|} \big[ \left(1 -2R_0\right) R_2 - R_1^2 \big]dt  - \dfrac{1}{|\alpha-\beta|} \, R_1 \, dW_1 \label{eq57}\\[2ex]
d R_2  = & \big[ \big(6 R_0 - 3 R_0^2 -2 \big) \, R_2 + 3 (1 - R_0) R_1^2 \big] \, dt + 2(1-R_0)\,R_1 \, d W_2 \label{eq58}
\end{align}
Taking the stochastic expectation on both sides of \eqref{eq55}--\eqref{eq58} and using the zero expectation property of It\^o stochastic integrals yields
%\begin{equation} \begin{array}{rcl}
%\dfrac{d E[\theta_1]}{dt} & = &  E\left[\dfrac{d\theta_1}{dt} \right] = \dfrac{2\beta}{|\alpha - \beta|} (1-2R_0) E[R_1] \\[2ex]
%\dfrac{d E[R_1]}{dt} & = & \left(-2+6R_0-3R_0^2 \right) E[R_1] \\[2ex]
%\dfrac{d E[\theta_2]}{dt} & = &  E\left[\dfrac{d\theta_2}{dt} \right] = \dfrac{2\beta}{|\alpha - \beta|} \big[-E[R_1^2] + (1-2R_0) E[R_2]\big]  \\[2ex]
%\dfrac{d E[R_2]}{dt} & = & 3\left(1-R_0\right) E[R_1^2] + \left(6R_0-3R_0^2-2\right) E[R_2]\\[2ex]
%\end{array} \label{eq56} \end{equation}
\begin{align}
\dfrac{d E[\theta_1]}{dt}  = &  E\left[\dfrac{d\theta_1}{dt} \right] = \dfrac{2\beta}{|\alpha - \beta|} (1-2R_0) E[R_1] \label{eq59}\\[2ex]
\dfrac{d E[R_1]}{dt}  = & \left(-2+6R_0-3R_0^2 \right) E[R_1] \label{eq60}\\[2ex]
\dfrac{d E[\theta_2]}{dt}  = &  E\left[\dfrac{d\theta_2}{dt} \right] = \dfrac{2\beta}{|\alpha - \beta|} \big[-E[R_1^2] + (1-2R_0) E[R_2]\big]  \label{eq61}\\[2ex]
\dfrac{d E[R_2]}{dt}  = & \left(6R_0-3R_0^2-2\right) E[R_2]  + 3\left(1-R_0\right) E[R_1^2] \label{eq62}
\end{align}
To close system \eqref{eq59}--\eqref{eq62}, an ODE for $E[R_1^2]$ is needing. Using It\^o formula the SDE $d(R_1)^2 = 2 R_1 dR_1 + (dR_1)^2$ is obtained, with $(d R_1)^2 = (1-R_0)^4 dt$ as a consequence of \eqref{eq56} and of It\^o lemma. After substitution and  taking the expectation value the closing equation is found
\begin{equation}
\dfrac{d E[R_1^2]}{dt} = \left(-4+12R_0-6R_0^2\right) E[R_1^2] + (1-R_0)^4 \label{eq63}
\end{equation}

Integrating \eqref{eq54} by separation of variables it is found that, as expected, $R_0(t)\rightarrow 0$ asymptotically for $t\rightarrow +\infty$. Taking this into account the asymptotic stationary solution of system \eqref{eq59}--\eqref{eq63} are easily found $E[R_1] =0$,  $E[R_1^2] = 1/4$, $E[R_2] = 3/8$, $E[d\theta_1/dt] = 0$, and $E[d \theta_2/dt] = \beta/(4|\alpha-\beta|)$. It follows that the expected angular frequency, amplitude and amplitude variance are
\begin{equation}
E\left[\dfrac{d \theta}{dt} \right] = 1 + \varepsilon^2\,\dfrac{\beta}{4|\alpha-\beta|}, \qquad E[R] = \dfrac{3}{8} \, \varepsilon^2, \qquad  \textrm{Var}[R] = \dfrac{1}{4}\,\varepsilon^2 \label{eq64}
\end{equation}
%while the expected amplitude is
%\begin{equation}
%E[R] = \dfrac{3}{8} \, \varepsilon^2 \label{eq65}
%\end{equation}
%with variance
%\begin{equation}
%\textrm{Var}[R] = \dfrac{1}{4}\,\varepsilon^2 \label{eq66}
%\end{equation}

\subsection{Phase and amplitude equations with Floquet's basis}

The Jacobian matrix evaluated over the limit cycle is
\begin{equation}
\bA(t) = \left[ \begin{array}{cc}
0 & -2\beta\\
0 & -2
\end{array} \right]  \label{eq65}
\end{equation}
with eigenvalues $\mu_1=0$, $\mu_2 = -2$. The associated eigenvectors are the Floquet's vectors $\bu_1(t) = [1,0]^T$ and $\bu_2(t)=[\beta,1]^T$, while inverting the matrix $\bU(t)=[\bu_1(t),\bu_2(t)]$ we find the Floquet's co--vectors $\bv_1(t)=[1,-\beta]$ and $\bv_2(t)=[0,1]$. Repeating the calculations of the previous section we find that the relation between the old and the new coordinates is $\varphi = (\alpha - \beta) \theta + \beta R$, $\rho = 1 +R$. The phase and amplitude equations in the new basis are
\begin{equation} \begin{array}{rcl}
d\theta & = & \left(1+\dfrac{\beta\,R^2}{|\alpha - \beta|} (2+R)\right) dt + \dfrac{\varepsilon}{|\alpha -\beta|} \, (1-R) \, dW_1\\[2ex]
d R & = & -\left( 2R+3R^2+R^3\right) dt + \varepsilon \, (1+R)^2 dW_2
\end{array} \label{eq66} \end{equation}
Conversely to the orthogonal case, the phase equation obtained using Floquet's basis does not contain a drift term linear in $R$, in accordance with theorem \ref{theorem3}. The new equations for the first terms of the asymptotic expansion are
\begin{align}
d \theta_0 = & \left[1+\frac{\beta}{|\alpha - \beta|} \left(2R_0^2 + R_0^3 \right) \right] dt \label{eq67} \\[2ex]
d R_0 = & -\left(2 R_0 + 3 R_0^2 + R_0^3 \right) dt \label{eq68}\\[2ex]
d \theta_1  = & \frac{\beta}{|\alpha - \beta|} \left(4 R_0 + 3 R_0^2 \right) R_1 \, dt + \frac{1}{|\alpha-\beta|} \big[(1+R_0) \, dW_1 - \beta (1+R_0)^2 dW_2 \big] \label{eq69} \\[1ex]
d R_1  = & -\left(2 + 6 R_0 + 3 R_0^2 \right) R_1 \, dt + (1+R_0)^2 \, d W_2 \label{eq70}\\[2ex]
\nonumber d \theta_2  = & \frac{\beta}{|\alpha - \beta|} \big[ \big(4 R_0 + 3 R_0^2 \big) R_2 + (2 + 3R_0) R_1^2 \big] dt \\
& + \frac{1}{|\alpha-\beta|} \big[R_1 \, dW_1 -2\beta(1+R_0)R_1 dW_2 \big] \label{eq71}\\[2ex]
d R_2  = & -\big[ \big(2 + 6 R_0 + 3 R_0^2 \big) R_2 + 3 (1 + R_0) R_1^2 \big] \, dt + 2(1+R_0)R_1 \, d W_2 \label{eq72}
\end{align}
The equations for the stochastic expectations read
\begin{align}
\dfrac{d E[\theta_1]}{dt}  = &  E\left[\dfrac{d\theta_1}{dt} \right] = \dfrac{\beta}{|\alpha - \beta|} (4R_0+3R_0^2) E[R_1] \label{eq73}\\[2ex]
\dfrac{d E[R_1]}{dt}  = & -\left(2+6R_0+3R_0^2 \right) E[R_1] \label{eq74}\\[2ex]
\dfrac{d E[\theta_2]}{dt}  = &  E\left[\dfrac{d\theta_2}{dt} \right] = \dfrac{\beta}{|\alpha - \beta|} \big[(4R_0+3R_0^2) E[R_2] + 2 E[R-1^2] + 3 R_0 E[R_1^2] \big]  \label{eq75}\\[2ex]
\dfrac{d E[R_2]}{dt}  = & -\left(2+6R_0+3R_0^2\right) E[R_2] - 3 (1+R_0) E[R_2]\label{eq76}\\[2ex]
\dfrac{d E[R_1^2]}{dt} = & -\left(4+12R_0+6R_0^2\right) E[R_1^2] + (1-R_0)^4 \label{eq77}
\end{align}
where equation \eqref{eq77} has been obtained by the same procedure described in the previous section. The asymptotic stationary solutions are $E[R_1]=0$, $E[R_1^2]=1/4$, $E[R_2]=-3/8$, $E[d\theta_1/dt]=0$, $E[d \theta_2/dt]=\beta/(2|\alpha-\beta|)$. Finally, the expected angular frequency, amplitude and amplitude variance are
expected angular frequency are
\begin{equation}
E\left[\dfrac{d \theta}{dt} \right] = 1 + \varepsilon^2\,\dfrac{\beta}{2|\alpha-\beta|}, \qquad E[R] = -\dfrac{3}{8} \, \varepsilon^2, \qquad \textrm{Var}[R] = \dfrac{1}{4}\,\varepsilon^2  \label{eq78}
\end{equation}

\begin{figure}
\centering
 \includegraphics[width=65mm]{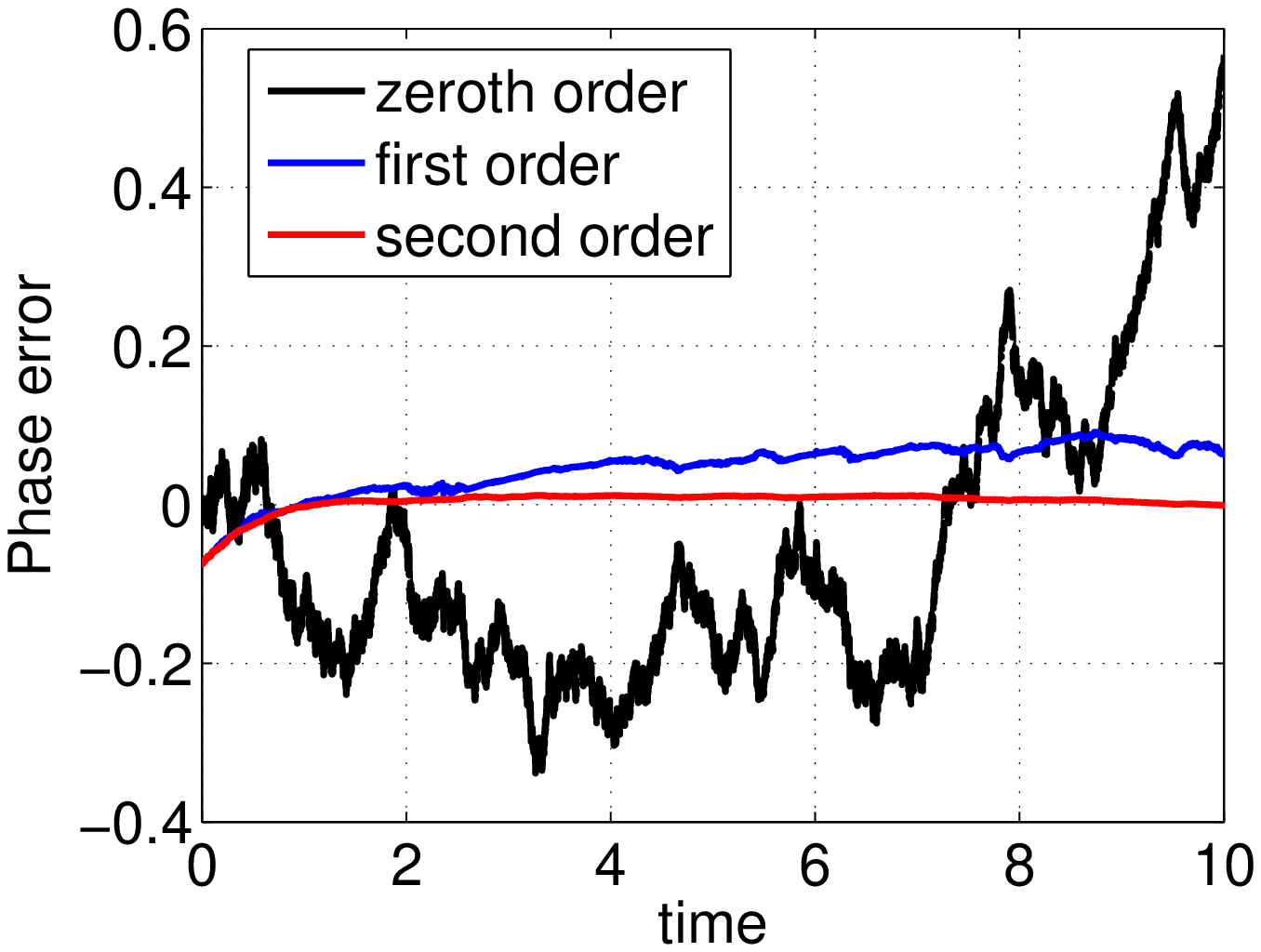}
 \includegraphics[width=65mm]{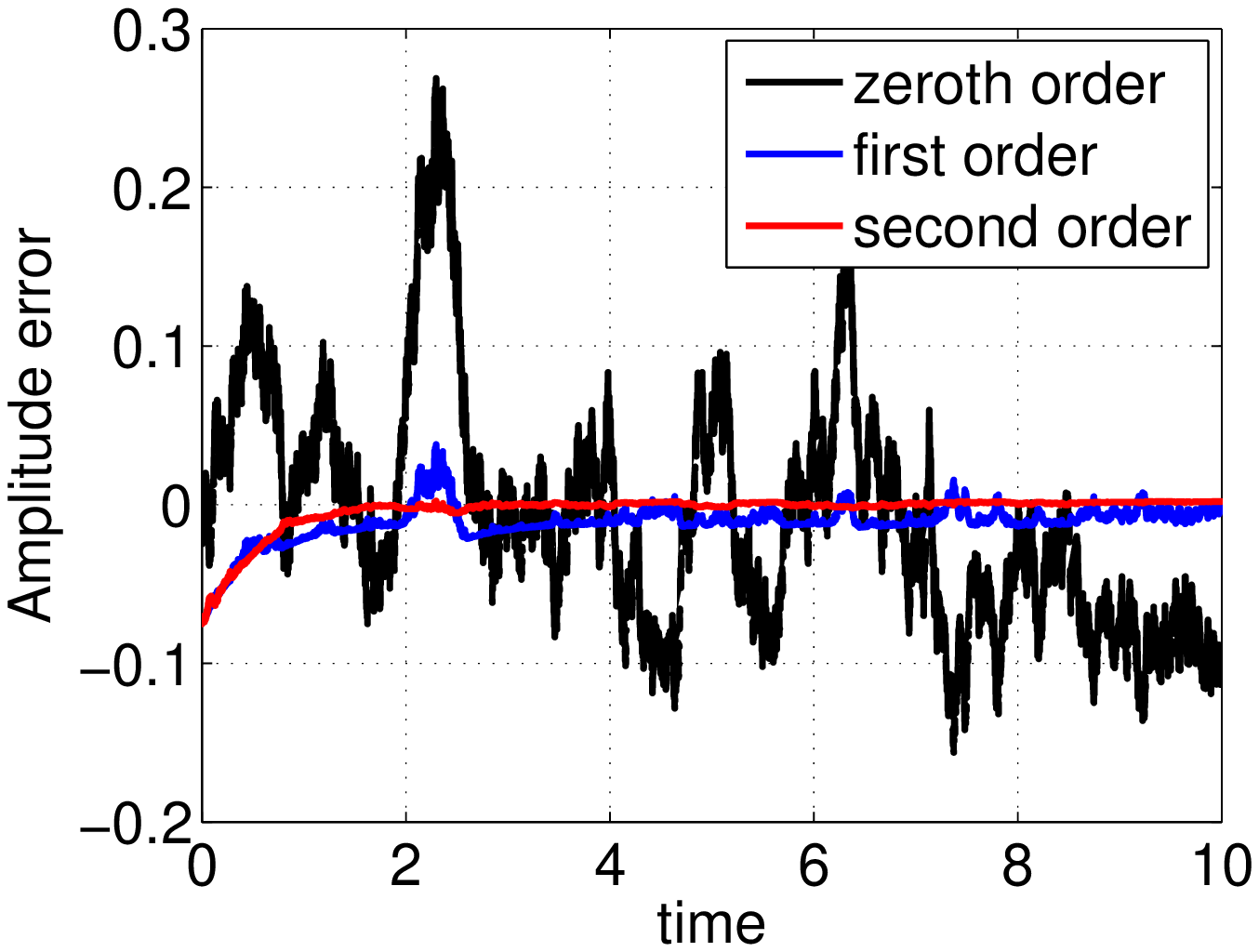}%
\caption{Accuracy of the asymptotic expansion: phase and amplitude error vs time for various order od approximation and a specific realization of the Brownian motion. Floquet's basis is used. Parameters are $\alpha = 5$, $\beta = 1$, and $\varepsilon = 0.15$ \label{figure1}}
\end{figure}

The theoretical predictions have been compared with Monte--Carlo simulations. The Stuart--Landau equation \eqref{eq50} has been integrated numerically using both Euler--Maruyama and Milstein integration schemes. Figure \ref{figure1} shows the difference versus time, between the numerical solution of \eqref{eq50} and the asymptotic expansion solution, for different order of approximation. The result has been obtained using Floquet's basis and is relative to a specific realization of the noise. It is indicative of the increasing level of accuracy of the expansion.

\begin{figure}
\centering
 \includegraphics[width=65mm]{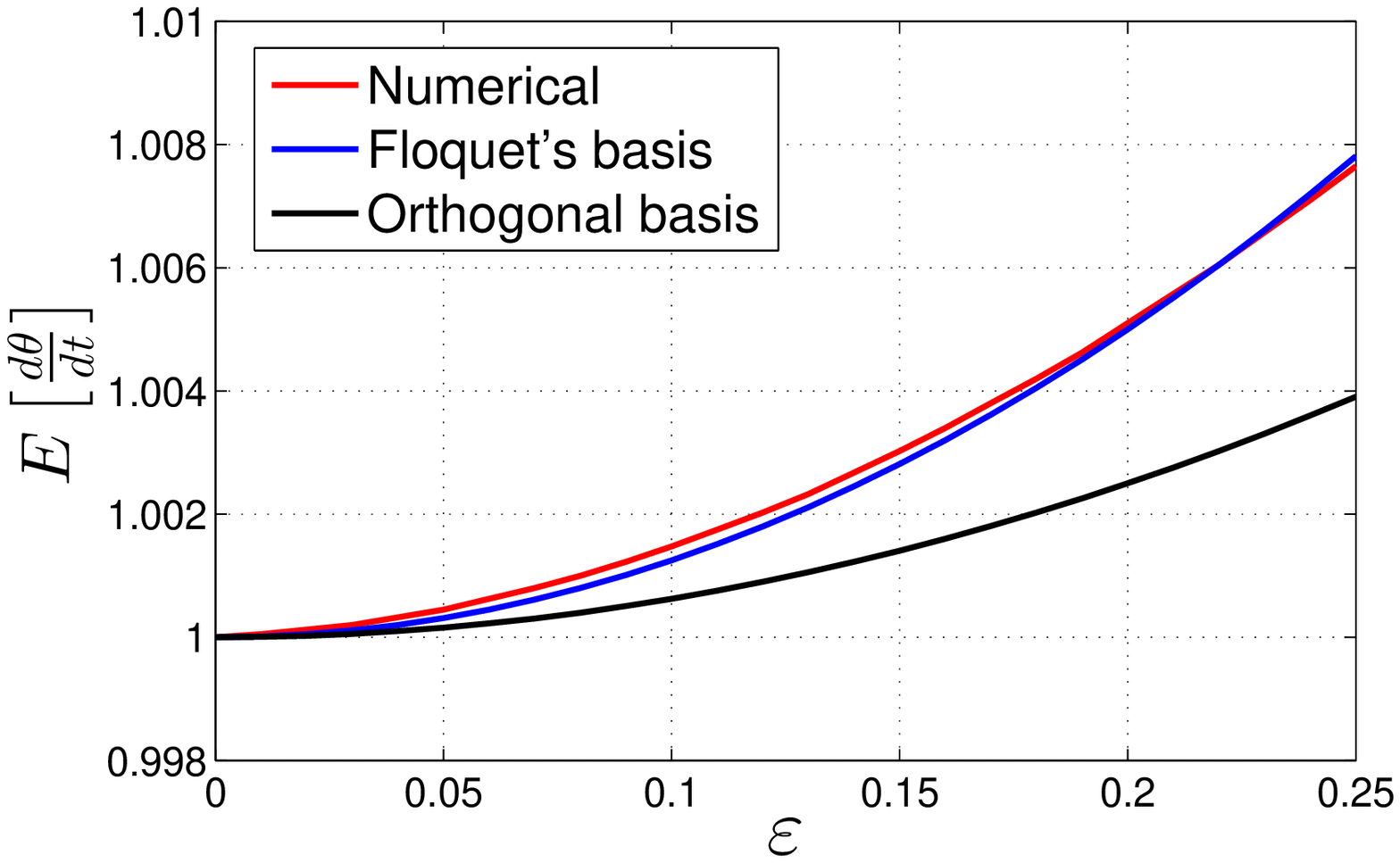}
 \includegraphics[width=65mm]{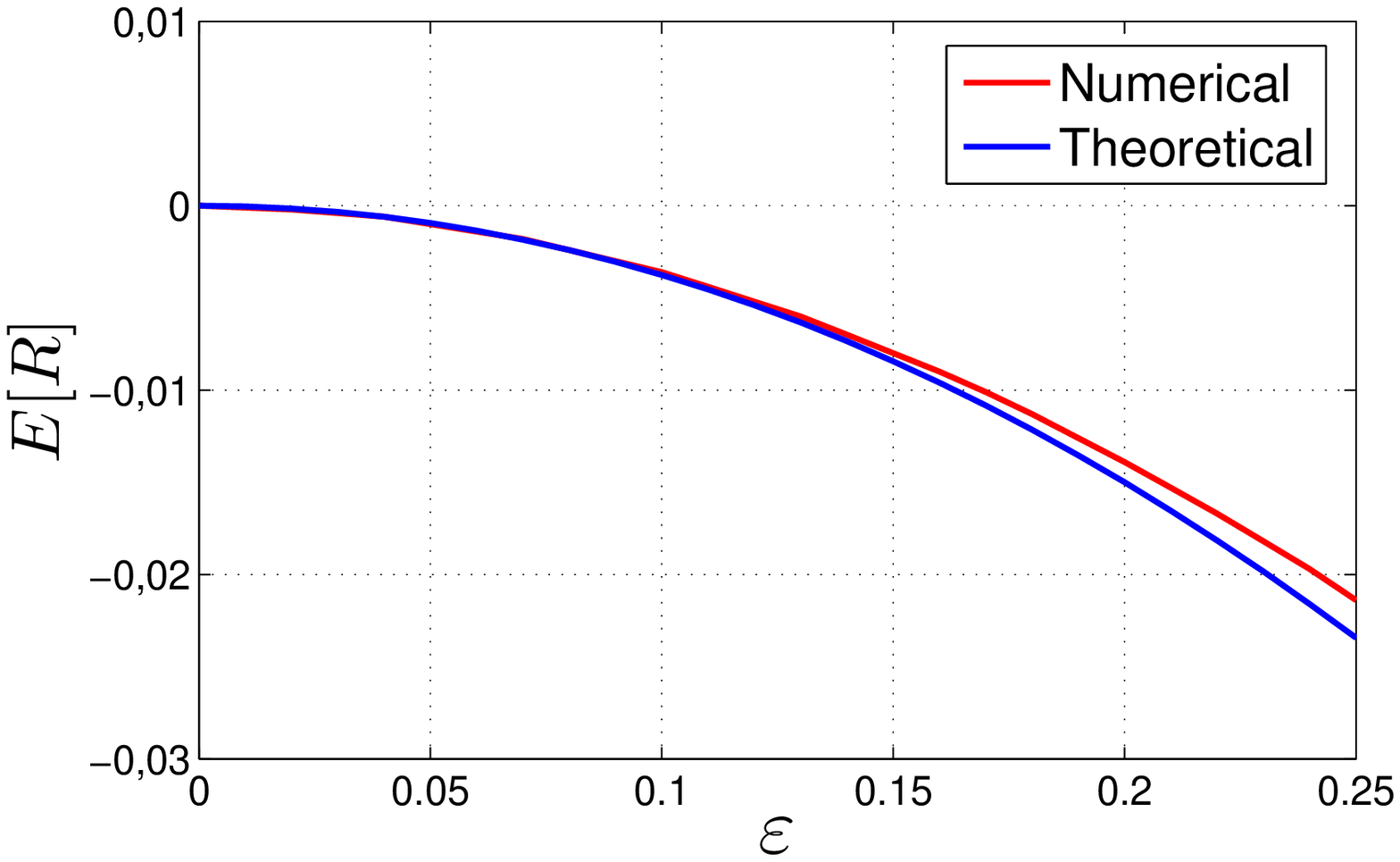}
\caption{Left: Expected normalized angular frequency vs noise intensity, for the Stuart--Landau oscillator. Right: Expected amplitude deviation vs noise intensity. Theoretical predictions are obtained using both the orthogonal basis and Floquet's basis.  The values of the parameters are $\alpha = 5$, $\beta = 1$. \label{figure2}}
 \end{figure}

Figure \ref{figure2} shows the expected amplitude and angular frequency versus the noise intensity, calculated using the asymptotic expansion method. The theoretical prediction for the expected amplitude obtained using orthogonal basis and Floquet's basis coincide. For the expected angular frequency, the Floquet's basis gives a more accurate result, as a consequence of the partial decoupling between phase and amplitude.

\begin{figure}
\centering
 \includegraphics[width=65mm]{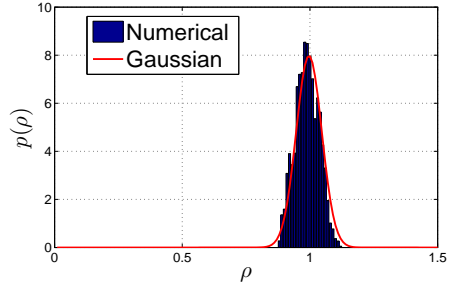}%
 \caption{Stationary amplitude distribution $p(\rho,t\rightarrow +\infty)$, for the Stuart--Landau oscillator \eqref{eq50}. A Gaussian distribution with mean $\mu=1+E[R]$ and variance $\sigma^2=\textrm{var}(R)$ is shown for comparison. Parameters are: $\alpha = 5$, $\beta=1$, $\varepsilon =0.15$.\label{figure3}}
 \end{figure}

Finally figure \ref{figure3} shows the stationary distribution for the probability density of the amplitude ($p(\rho,t)$ for $t\rightarrow \infty$). The experimental probability to find the amplitude in the interval $\rho + d \rho$ is evaluated as the fraction of simulation time spent in the interval. A Gaussian distribution with mean $1+E[R]$ and variance $\textrm{var}(R)$ is also shown for comparison.

\section{Conclusions}\label{conclusions}

A novel description for nonlinear oscillators subject to white Gaussian noise and described by It\^o stochastic differential equations, is given. The dynamics is described in terms of a phase and an amplitude deviation variables. The phase function defines the projection of the stochastic orbit onto a reference trajectory, i.e. a limit cycle of the unperturbed system, while the amplitude represents the deviation from the reference orbit.

The specific equations that are obtained depend upon the choice of a particular basis vectors. Two main basis have been considered: orthogonal basis and  Floquet's basis. Although the orthogonal basis is conceptually simpler, it has been shown that Floquet's basis allows a partial decoupling between the amplitude and the phase equations. This permits to obtain more accurate estimation of the expected angular frequency, and to derive simplified phase reduced models.

In the weak noise limit, the amplitude and phase equation can be conveniently solved with the help of asymptotic expansions. The method recast the equations in the form of a sequence of time dependent Ornstein--Uhlenbeck processes, that can be solved iteratively. In alternative, useful information concerning the expectation values of amplitude and angular frequency can be obtained using It\^o calculus. Taking stochastic expectations and using the zero expectation property of It\^o integrals a system of ordinary differential equations for expectation values can be obtained. Due to the fact that Ornstein--Uhlenbeck processes are linear stochastic differential equations, the system is closed and can be solved analytically thus allowing a complete characterization of the stochastic process.

\section*{Acknowledgements}
This work was partially supported by the Ministry of Foreign Affairs (Italy) \emph{``Con il contributo del Ministero degli Affari Esteri, Direzione Generale per la Promozione del Sistema Paese.''}

\section{Appendix A: Floquet theory basics} \label{floquet theory}
Consider the $n$--dimensional homogeneous linear system of ODEs
\begin{equation}
\dfrac{d\by(t)}{dt} = \bA(t) \by(t) \label{app-eq1}
\end{equation}
where $\bA(t)$ is a $n\times n$ dimensional matrix with periodic entries. Let $\by_1(t),\ldots,\by_n(t)$ be $n$ linearly independent solutions of \eqref{app-eq1}. Then
\begin{itemize}
\item $\Phi(t) = [\by_1(t),\ldots,\by_n(t)]$ is called a fundamental matrix. If $\Phi(0) = \bI_n$ then $\Phi(t)$ is called the principal fundamental matrix of the state transition matrix.
\item Any solution of \eqref{app-eq1} can be written in the form $\by(t) = \Phi(t) \by(0)$.
\item The fundamental matrix is not unique, but they are all similar, that is, if $\Phi(t)$ and $\widetilde \Phi(t)$ are fundamental matrices, then a constant matrix $\bC$ exists such that $\widetilde \Phi(t) = \Phi(t) \bC$.
\item Let $A(t)=A(t+T)$, then $\Phi(t+T)$ is also a fundamental matrix, and $\Phi(t+T)=\Phi(t) \bD$. The eigenvalues of the constant, non singular matrix $\bD$, $\mu_1,\ldots,\mu_n$ are called Floquet's characteristic multipliers. They are related to the Floquet's characteristic exponents, $\nu_1,\ldots,\nu_n$ by the formula $\mu_i = e^{\nu_i T}$.
\end{itemize}
\begin{theorem}
If $\mu$ is a characteristic multiplier of system \eqref{app-eq1}, then a non trivial solution $\by_s(t)$ exists satisfying
\begin{equation}
\by_s(t+T) = \mu \, \by_s(t) \label{app-eq2}
\end{equation}
Viceversa, if a nontrivial solution $\by_s(t)$ of \eqref{app-eq1} satisfies \eqref{app-eq2}, then $\mu$ is a characteristic multiplier and $\by_s(0)$ is the corresponding eigenvector.
\end{theorem}
\begin{theorem}[Floquet 1883]
The fundamental matrix $\Phi(t)$ of system \eqref{app-eq1} can be written in the form
\begin{equation}
\Phi(t) = \bP(t) \, e^{\bD\,t} \, \bS_0
\end{equation}
where $\bD = \diag[\nu_1,\ldots,\nu_n]$, $\bP(t)$ is a $T$--periodic regular matrix and $\bS_0=\bP^{-1}(0)$
\end{theorem}
\begin{theorem}
If $\Phi(t)$ is a fundamental matrix of \eqref{app-eq1}, then $\Psi(t) = \left[\Phi^{-1}(t)\right]^T$ is a fundamental matrix for the adjoint problem
\begin{equation}
\dfrac{d\bz(t)}{dt} = -\bA^T(t) \, \bz(t) \label{app-eq3}
\end{equation}
\end{theorem}
Therefore, denoting by $\bp_i(t)$  the $i^{th}$ column of $\bP(t)$ and by $\bs_i(t)$ the $i^{th}$ row of $\bP^{-1}(t)$, if $\by_s(t) = \bp_i(t) \, e^{\nu_i t}$ is a solution of \eqref{app-eq1}, then $\bz_s(t) = \bs_i(t) \, e^{-\nu_i t}$ is a solution of \eqref{app-eq3}.

\section*{Appendix B: some details on It\^o calculus}

Stochastic processes are nowhere differentiable. As a consequence the SDEs \eqref{eq1} should be interpreted as a shorthanded notation for the integral equation
\begin{equation}
\bX_t = \bX_0 + \int_0^t \ba(\bX_s) \, ds + \int_0^t \bB(\bX_s) \, d\bW_s \label{app-eq4}
\end{equation}
Depending on the how the second integral on the right hand side is defined, different interpretations are possible. The most popular interpretations are Stratonovich and It\^o.
\begin{itemize}
\item Stratonovich integral. We adopt the standard notation $\f(\bX_t) \circ d\bW_t$ to denote Stratonovich interpretation. In this view the functions are evaluated at the middle point of each time interval, that is
\begin{equation}
\int_0^T \f(\bX_t) \circ d\bW_t = \textrm{ms}-\hspace{-3mm}\lim_{n\rightarrow +\infty} \sum_{j=1}^n \f\left(\dfrac{\bX_{t_j} + \bX_{t_{j-1}}}{2}\right) \left( \bW_{t_j} - \bW_{t_{j-1}}\right)
\end{equation}
    where $\textrm{ms-lim}$ denotes the mean square limit.
\item It\^o stochastic integral. In It\^o interpretation the functions are evaluated at the beginning of each interval
\begin{equation}
\int_0^T \f(\bX_t) d\bW_t = \textrm{ms}-\hspace{-3mm}\lim_{n\rightarrow +\infty} \sum_{j=1}^n \f\left(\bX_{t_{j-1}}\right) \left( \bW_{t_j} - \bW_{t_{j-1}}\right)
\end{equation}
\item It\^o lemma states that
\begin{equation}
dt^2 = dt \, dW_i = dW_i \, dt=0, \qquad dW_i \, dW_j = \delta_{ij} dt
\end{equation}
\item It\^o SDEs do not follow the traditional calculus rule for change of variables. It\^o formula must be used instead. Let $\bX_t$ be an It\^o process solution of the SDE \eqref{eq1}, and let $\bg:\R^n \mapsto \R^n$ be a twice differentiable function, then $\bg(\bX_t)$ is again an It\^o process and its components are given by
\begin{equation}
d g_k = \sum_{i=1}^n \dfrac{\partial g_k}{\partial x_i} \, dX_i + \dfrac{1}{2} \sum_{i,j=1}^n \dfrac{\partial^2 g_k}{\partial x_i \partial x_j} \, dX_i dX_j
\end{equation}
\item Since in It\^o interpretations stochastic processes and noise increments are independent, It\^o integrals have zero expectation value
\begin{equation}
E\left[\int_T^S f(t) \,dW_t \right] = 0
\end{equation}
\end{itemize}

\section*{References}
  \bibliographystyle{elsarticle-num}
  \bibliography{bonnin_phase_amplitude}

\end{document}